\documentclass[conference]{IEEEtran}
\usepackage{color}
\usepackage{graphicx}
\usepackage{epstopdf}
\usepackage{amsmath}
\usepackage{amssymb}
\usepackage{hyperref}
\usepackage[english]{babel}
\usepackage{cite}
\usepackage{rotfloat}
\usepackage{mathtools}
\usepackage{amsmath}
\usepackage{makecell}
\usepackage{algorithm,algorithmic}
\usepackage{multirow}
\usepackage{subfigure}
\usepackage{booktabs}
\usepackage{colortbl}
\usepackage{multirow}% http://ctan.org/pkg/multirow
\usepackage{hhline}% http://ctan.org/pkg/hhline
\usepackage{stfloats}% <-- added
\usepackage{multicol}
\usepackage{bbm}
\usepackage{cases}
\usepackage{setspace}
\allowdisplaybreaks
\usepackage[status=draft,layout=margin]{fixme}
\usepackage[justification=centering,font=small,skip=0pt]{caption}

\newsavebox{\foobox}

\newcommand{\setA}{\mathcal{A}}

\definecolor{kugray5}{RGB}{224,224,224}

\usepackage[normalem]{ulem}
\newcommand\rsout{\bgroup\markoverwith
	{\textcolor{red}{\rule[0.5ex]{2pt}{0.8pt}}}\ULon}



\makeatletter
\newcommand{\ALOOP}[1]{\ALC@it\algorithmicloop\ #1%
	\begin{ALC@loop}}
	\newcommand{\ENDALOOP}{\end{ALC@loop}\ALC@it\algorithmicendloop}

\makeatother

\usepackage{etoolbox}
\let\mybibitem\bibitem
\renewcommand{\bibitem}[1]{%
	\ifstrequal{#1}{nature}
	{\color{blue}\mybibitem{#1}}
	{\color{black}\mybibitem{#1}}%
}

\graphicspath{ {Figures/} }

\DeclareCaptionLabelSeparator{periodspace}{.\quad}

\renewcommand{\figurename}{Fig.}
\addto\captionsenglish{\renewcommand{\figurename}{Fig.}}
\newcommand\nbthis{\addtocounter{equation}{1}\tag{\theequation}}

\newcommand{\norm}[1]{\left\lVert#1\right\rVert} % ||.||
\newcommand{\normshort}[1]{\lVert#1\rVert} % ||.||
\newcommand{\abs}[1]{\left|#1\right|} % ||
\newcommand{\absshort}[1]{\lvert#1\rvert} % ||

\newcommand{\diagshort}[1]{\mathrm{diag}\{#1\}} % ||
\newcommand{\blkdiagshort}[1]{\mathrm{blockdiag}\{#1\}} % ||

\newcommand{\re}[1]{\mathfrak{R}{\left(#1\right)}}

\newcommand{\meanshort}[1]{\mathbb{E} \{#1\}}

\newcommand{\mQ}{\textbf{\textit{Q}}}

\newcommand{\mH}{\textbf{\textit{H}}}

\newcommand{\mI}{\textbf{\textit{I}}}

\newcommand{\setC}{\mathbb{C}} 

\newcommand{\vx}{\textbf{\textit{x}}}

\newcommand{\vn}{\textbf{\textit{n}}}

\newcommand{\vh}{\textbf{\textit{h}}}

\newcommand{\vw}{\textbf{\textit{w}}}

\newcommand{\vt}{\textbf{\textit{t}}}

\newcommand{\va}{\boldsymbol{\alpha}}

\newcommand{\bPhi}{\boldsymbol{\Phi}}
\newcommand{\bUpsilon}{\boldsymbol{\Upsilon}}

\newcommand{\bPsi}{\boldsymbol{\Psi}}

\newcommand{\an}{\alpha_n}

\newcommand{\Na}{N_{\mathrm{a}}}

\newcommand{\hd}{\vh_{0,k}}

\newcommand{\hrt}{\mH_{1}} 
 
\newcommand{\hrr}{\vh_{2,k}}

\newcommand{\ptmax}{p^{\mathrm{max}}_{\mathrm{BS}}} 
\newcommand{\prismax}{p^{\mathrm{max}}_{\mathrm{RIS}}} 
\newcommand{\pris}{p_{\mathrm{RIS}}} 
\newcommand{\pbs}{p_{\mathrm{BS}}}

\newcommand{\ba}{\{\an\}}
\newcommand{\bXi}{\boldsymbol{\Xi}} 
\newcommand{\bW}{\{\vw_k\}} 

\begin{document}
	\title{Hybrid Active-Passive Reconfigurable Intelligent Surface-Assisted Multi-User MISO Systems}
	\author{\IEEEauthorblockN{Nhan~T.~Nguyen\IEEEauthorrefmark{1}, V.-Dinh~Nguyen\IEEEauthorrefmark{2}, Qingqing~Wu\IEEEauthorrefmark{3}, Antti~Tölli\IEEEauthorrefmark{1},  Symeon~Chatzinotas\IEEEauthorrefmark{2}, and Markku~Juntti\IEEEauthorrefmark{1}}
		%% Affiliation
		\IEEEauthorblockA{\IEEEauthorrefmark{1}Centre for Wireless Communications, University of Oulu, P.O.Box 4500, FI-90014, Finland}
		\IEEEauthorblockA{\IEEEauthorrefmark{2}Interdisciplinary Centre for Security, Reliability and Trust (SnT), University of Luxembourg, L-1855 Luxembourg}
		\IEEEauthorblockA{\IEEEauthorrefmark{3}State Key Laboratory of Internet of Things
			for Smart City, University of Macau, Macau 999078, China}
		\{nhan.nguyen, antti.tolli, markku.juntti\}@oulu.fi; \{dinh.nguyen, symeon.chatzinotas\}@uni.lu; qingqingwu@um.edu.mo
		\vspace{-10pt}}
	
	\maketitle
	
	\begin{abstract}
		
		We consider a multi-user multiple-input single-output (MISO) communications system which is assisted by a hybrid active-passive reconfigurable intelligent surface (RIS). Unlike  conventional passive RISs, hybrid RIS is equipped with a few active elements with the ability to reflect and amplify incident signals to significantly improve the system performance. Towards a fairness-oriented design, we maximize the minimum rate among all users through jointly optimizing the transmit beamforming vectors and RIS reflecting/amplifying coefficients. Combining tools from block coordinate ascent and successive convex approximation, the challenging nonconvex problem is efficiently solved by a low-complexity iterative algorithm. The numerical results show that a hybrid RIS with $4$ active elements out of a total of 50 elements with a power budget of $-1$ dBm offers an improvement of up to $80\%$ to the considered system, while that achieved by a fully passive RIS is only $27\%$.
		
	\end{abstract}
	
	\begin{IEEEkeywords}
		Hybrid active-passive RIS, multi-user MISO, beamforming, RIS semi-passive beamforming, successive convex approximation.
	\end{IEEEkeywords}
	\IEEEpeerreviewmaketitle
	
	\section{Introduction}
	
	Reconfigurable intelligent reflecting surfaces (RISs) have been advocated as a cost and energy-efficient solution to improve the performance of wireless communications systems \cite{huang2019reconfigurable, wu2019intelligent}. The reflecting elements of RISs can be configured to not only improve the received signal power but also mitigate interference in multi-user wireless systems \cite{wu2019towards}. The performance improvement of RISs in multi-user multiple-input multiple-output (MISO) systems are investigated in \cite{hu2020reconfigurable, di2020hybrid, kammoun2020asymptotic, zhang2021reconfigurable, li2020interference, gan2021ris, di2020practical, huang2020reconfigurable, ma2021joint, huang2018energy}. Through the joint optimization of the transmit beamforming/power at the base station (BS) and the reflecting coefficients of the RIS, it is shown in  \cite{hu2020reconfigurable, kammoun2020asymptotic, gan2021ris, di2020practical, ma2021joint, huang2020reconfigurable} that the system achievable sum-rate and/or fairness can be significantly improved thanks to the deployment of RISs. Li \textit{et al.} \cite{li2020interference} exploited the concept of constructive interference and proposed an efficient design of RIS coefficients to minimize the symbol error rate. Whereas, the works in \cite{gan2021ris} and \cite{di2020practical} focused on practical scenarios, where only statistical channel information and/or low-resolution phase shifts at the RIS are available for the joint design. To overcome the challenging nonconvexity and strongly coupled variables in the design, Huang \textit{et al.} \cite{huang2020reconfigurable} developed a deep reinforcement learning-based algorithm to simultaneously obtain the transmit beamformers and RIS phase shifts. In particular, Zhang \textit{et al.} \cite{zhang2021reconfigurable} derived an expression of the system asymptotic capacity and determined the required number of RIS elements to meet a predefined quality of service (QoS).
	
	All the above works considered the passive RIS, in which reflecting elements are unable to amplify  incident signals, and, thus, a large number of elements are required to compensate for the severe cascaded path loss \cite{wu2019intelligent}. Recently, hybrid active-passive RIS architectures have been introduced \cite{ taha2019deep, alexandropoulos2020hardware, nguyen2021spectral,nguyen2021hybrid,nguyen2021hybrid_mag} to overcome the inherent limitation of  passive RISs, especially in harsh transmission scenarios such as in low signal-to-noise ratio (SNR) regime and/or severe path loss. The key idea of the hybrid RIS is to add a few active elements to the conventional passive RIS, allowing them to reflect and amplify incident signals simultaneously. As a result, the hybrid RIS can reduce effects of the double path loss and significantly improve the system performance in terms of spectral efficiency \cite{nguyen2021spectral, nguyen2022downlink, zeng2021active}, secrecy rate \cite{9598322}, harvested energy \cite{9653007}, and reliability \cite{DBLP:journals/corr/abs-2111-08355}. These advantages are also reaped by fully active RISs \cite{long2021active, khoshafa2021active, chen2022active}, which, however, come at high cost of power consumption and hardware design, compared to the hybrid RIS with only a few active elements. Moreover, active elements with RF chains for processing of incident signals and channel estimation at the RIS are presented in \cite{taha2019deep, alexandropoulos2020hardware,schroeder2020passive}.
	
	In this work, we deploy the hybrid active-passive RIS to assist the multi-user MISO system. Thanks to the signal amplification, the hybrid RIS with a small-to-moderate size can efficiently compensate for the severe path loss and blockage in the communication links, especially in complex urban scenarios. Focusing on the fairness design, our goal is to maximize the minimum rate among all user equipments (UEs) by jointly optimizing the transmit beamformers and RIS amplifying/reflecting coefficients. To the best of the authors' knowledge, it has not been investigated in the literature. The problem is practically appealing but particularly more challenging, compared to those with conventional passive RISs, due to additional power constraints and amplified noise/interference caused by RIS active elements. To effectively solve the challenging optimization problem, we propose an efficient algorithm based on the block coordinate ascent (BCA) and successive approximation (SCA). Specifically, the original problem is first decomposed into two subproblems, which are then alternately solved by the SCA technique. Finally, the proposed design is evaluated by numerical results, which show that with the same total power budget, the hybrid RIS offers remarkable performance improvement compared to the systems without RIS and that with the conventional passive RIS, especially when the transmit power at the BS is limited.%\footnote{In this paper since the considered problem is maximization, the proposed iterative algorithm is referred to as block coordinate ascent.} 
	
	\section{System Model and Problem Formulation}
	
	\subsection{System Model}
	\label{sec_system_model}
	We consider a downlink wireless network where a multiple-antenna BS serves $K$ single-antenna UEs. The communication between BS and UEs is assisted by a hybrid active-passive RIS installed on the building facade. RIS is equipped with $N$ elements, out of which, $\Na$ ($\Na \ll N$) elements are activated. The positions of active elements are predetermined in $\setA \subset \{1,2,\ldots,N \}$ with $\abs{\setA}=\Na$. The RIS's active elements can potentially be realized by low-power reflection amplifiers \cite{landsberg2017low}. We refer readers to \cite{long2021active, nguyen2021hybrid, landsberg2017low} for more details on the reflection amplifier-based active RIS. It is seen that the fully passive RIS (i.e., with $\Na=0$) is just a special case of the hybrid RIS. Therefore, in this work, we will use the general term ``\textit{RIS}" for the discussion in the system model, while specific terms ``\textit{passive RIS}" or ``\textit{hybrid RIS}" are used in comparisons. 
	
	Let $\an$ denote the coefficient associated with the $n$th element of the RIS. We can express $\an$ as $\an = \abs{\an} e^{j \theta_n}$, where $\theta_n \in [0, 2\pi)$ represents the phase shift, $\abs{\an} \in [0,1]$ for $n \notin \setA$, and $\abs{\an} \leq a_{\mathrm{max}}$ for $n \in \setA$. Here, $a_{\mathrm{max}}$ is the maximum power amplification gain that the active load can provide, which is up to $40$ dB if active elements are realized by reflection amplifiers \cite{long2021active, landsberg2017low}. We note here that to mitigate interference in multi-user systems, the reflection amplitude of passive RIS elements may not necessarily be unity \cite{wu2019towards}. Let  $\bUpsilon \triangleq \text{diag} \{ \alpha_1, \ldots, \alpha_{N} \} \in \setC^{N \times N}$ be the diagonal matrix of the RIS coefficients. For ease of exposition in the following analysis, we define an additive decomposition $\bUpsilon = \bPhi + \bPsi$, where $\bPsi = \mathbbm{1}^{\setA}_N \circ \bUpsilon$ and $\bPhi = \left(\mI_{N} - \mathbbm{1}^{\setA}_N\right) \circ \bUpsilon$ contain the active and passive coefficients, respectively. Here, $\mathbbm{1}^{\setA}_N$ is an $N \times N$ diagonal matrix whose non-zero elements are all unity and have positions determined by $\setA$, and $\circ$ represents a Hadamard product.
	
	Let $\hd^H \in \setC^{1 \times N_t}$, $\hrt \in \setC^{N \times N_t}$, and $\hrr^H \in \setC^{1 \times N}$ denote the channels between  BS and UE $k$, between  BS and  RIS, and between  RIS and UE $k$, respectively. The effective channel between  BS and UE $k$ can be expressed as $\vh_{k}^H = \hd^H + \hrr^H \bUpsilon \hrt$. Denote by $s_k$ with $\meanshort{\abs{s_{k}}^2}=1$ and $\vw_k \in \setC^{N_t \times 1}$ the transmitted symbol and the beamforming vector intended for UE $k$, respectively. The transmitted signal from the BS can be given as $\vx = \sum_{k=1}^{K} \vw_k s_{k} \in \setC^{N_t \times 1}$. Thus, the total transmit power at the BS is $\pbs = \sum_{k=1}^{K} \norm{\vw_k}^2 \leq \ptmax$, where $\ptmax$ is the maximum transmit power of the BS. The received signal at UE $k$ can be given as
	\begin{align*}
		y_k &= \vh_{k}^H \vw_k s_{k} + \sum\nolimits_{j \neq k} \vh_k^H \vw_j s_j + n_k, \nbthis \label{eq_signal_model}
	\end{align*}
	where $n_k = \hrr^H \bPsi \vn_{\mathrm{r}} + n_{\mathrm{u}}$ is the aggregated noise at UE $k$, with $n_{\mathrm{u}} \sim \mathcal{CN}(0,\sigma_{\mathrm{u}}^2)$ being the additive white Gaussian noise (AWGN) at UE $k$; and $\vn_{\mathrm{r}} \sim \mathcal{CN} (\boldsymbol{0}, \mathbbm{1}^{\setA}_N \circ \sigma_{\mathrm{r}}^2  \mI_N)$ is the total effective noise including self-interference and AWGN noise caused by  RIS active elements operating in full-duplex mode \cite{nguyen2021hybrid}.% Let $\sigma_{\mathrm{r}}^2 = \sigma_{\mathrm{SI,r}}^2 + \sigma_{0,\mathrm{r}}^2$. Then, we have $\hrr^H \bPsi (\vz_{\mathrm{RSI,r}} + \vn_{0,\mathrm{r}}) \sim \mathcal{CN}(0,\sigma_{\mathrm{r}}^2 \normshort{\hrr^H \bPsi}^2)$ representing the effective noise at the UE received from the RIS. 
	
	\subsection{Problem Formulation}

	From \eqref{eq_signal_model}, the achievable rate of UE $k$ (in nats/s/Hz) can be expressed as
	\begin{align*}
		R_k = \log \left( 1 + \frac{\absshort{\vh_{k}^H \vw_k}^2}{ \sum\nolimits_{j \neq k} \absshort{\vh_k^H \vw_j}^2 + \sigma_{\mathrm{r}}^2 \normshort{\hrr^H \bPsi}^2 + \sigma_{\mathrm{u}}^2} \right). \nbthis \label{eq_rate}
	\end{align*}
	Let $\pris$ denote the transmit power of active elements of the RIS. It can be expressed as $\pris = \meanshort{\norm{\bPsi \left( \hrt \vx + \vn_{\mathrm{r}} \right) }^2} \overset{(a)}{=} \sum_{n \in \setA} \abs{\an}^2 \xi_n$, where $\xi_n \triangleq \sigma_{\mathrm{r}}^2 +  \norm{\vh_{1,n}}^2 \sum_{k=1}^{K} \norm{\vw_k}^2$; $\vh_{1,n}$ denotes the $n$th row of $\hrt$, and equality $(a)$ follows the diagonal structure of $\bPsi$ whose non-zero elements are in $\setA$. The total transmit power at the RIS is constrained as $\pris \leq \prismax$, where $\prismax$ is the power budget.%, we define
%	\begin{align*}
%		\xi_n \triangleq \sigma_{\mathrm{r}}^2 +   \lvert h_{1,n} \rvert^2. \nbthis \label{def_xi}
%	\end{align*}

 %\red{$\pris = \meanshort{\normshort{\bPsi (\vn_{\mathrm{r}} +   \hrt \vx) }^2} = \sum_{n \in \setA} \abs{\an}^2 \xi_n$, where $\xi_n = \sigma_{\mathrm{r}}^2 + \sum_{k=1}^K   \abs{h_{1,n}}^2$, with $h_{1,n}$ being the $n$th element of $\hrt$.} 
 
 We aim to maximize the minimum rate among all UEs through jointly optimizing the transmit beamformers and  RIS coefficients, which can be mathematically formulated as
	\begin{subequations}
		\label{prob_0}
		\begin{align}
			\underset{\substack{ \bW,\ba}}{\textrm{maximize}} \quad & \underset{k}{\textrm{min}} \{R_k\} \nbthis \label{opt_hyb_obj_SE} \\
			\textrm{subject to} \quad
			&0 \leq \sum\nolimits_{k=1}^{K} \norm{\vw_k}^2 \leq \ptmax, \nbthis \label{cons_tx_power} \\
			&0 \leq \theta_n \leq 2\pi, \ \forall n, \nbthis \label{cons_phase} \\
			&\abs{\alpha_n} \leq 1, \ \forall n \notin \setA, \nbthis \label{cons_passive_modul} \\
			&\abs{\alpha_n} \leq a_{\mathrm{max}}, \ \forall n \in \setA,  \nbthis \label{cons_active_modul} \\
			&\sum\nolimits_{n \in \setA} \abs{\an}^2 \xi_n \leq \prismax,  \nbthis \label{cons_RIS_power}
		\end{align}
	\end{subequations}
	where \eqref{cons_phase}--\eqref{cons_RIS_power} are constrains of the hybrid RIS. Note in \eqref{cons_active_modul} that is only active elements ($n \in \setA$) can amplify the signals with amplification gains restricted by $a_{\mathrm{max}}$ \cite{long2021active}. It is clear that the objective function is non-concave and non-smooth, resulting in a non-convexity of problem \eqref{prob_0}. 
	
	\section{Proposed Design}
	The strong coupling between $\bW$ and $\ba$  in the rate function makes problem \eqref{prob_0} difficult to solve. A direct application of SCA comes at a cost of high computational complexity. In what follows, we first transform problem \eqref{prob_0} into  a more tractable form as \eqref{prob_0} as
	\begin{subequations}
		\label{prob_1}
		\begin{align}
			\underset{\substack{\tau,  \bW,\ba}}{\textrm{maximize}} \quad & \tau \nbthis \label{opt_hyb_obj_SE_1} \\
			\textrm{subject to} \quad 
			&R_{k}  \geq \tau,\ \forall k, \nbthis \label{cons_min_rate} \\
			&\eqref{cons_tx_power}-\eqref{cons_RIS_power},
		\end{align}
	\end{subequations}
	where $\tau$ is an auxiliary variable.
	By utilizing the BCA approach, we decouple \eqref{prob_1} into two sub-problems with respect to $\bW$ and $\ba$, each of which is efficiently solved by the SCA method.
	
	\subsection{Transmit Beamforming Design}
	For given $\ba$, the optimal beamformers $\bW$ at the BS can be found by solving the following problem:
	\begin{align}
		\label{problem_power}
		\underset{\tau,\bW}{\textrm{maximize}}\ \tau,\ \textrm{subject to}\ \eqref{cons_min_rate}, \eqref{cons_tx_power}, \eqref{cons_RIS_power}.
	\end{align}
	where \eqref{cons_tx_power} and \eqref{cons_RIS_power} are convex. To convexify \eqref{cons_min_rate}, we introduce slack variables $\{\gamma_k\}$ to express it equivalently as
	\begin{subequations}
		\begin{align*}
			\log (1 + \gamma_{k}) &\geq \tau,\ \forall k, \nbthis \label{cons_min_rate_BF1} \\
			\frac{\absshort{\vh_{k}^H \vw_k}^2}{ \sum\nolimits_{j \neq k} \absshort{\vh_k^H \vw_j}^2 + \sigma_k^2} &\geq \gamma_{k},\ \forall k, \nbthis \label{cons_min_rate_BF2}
		\end{align*}
	\end{subequations}
	where $\sigma_k^2 \triangleq \sigma_{\mathrm{r}}^2 \normshort{\hrr^H \bPsi}^2 + \sigma_{\mathrm{u}}^2$ is a constant with respect to $\bW$. Let us define $\vw \triangleq [\vw_1^T, \ldots, \vw_K^T]^T \in \setC^{KN_t \times 1}$, $\tilde{\mH}_k \triangleq \vh_k \vh_k^H \in \setC^{N_t \times N_t}$, and
	\begin{align*}
	    \hat{\mH}_k \triangleq \blkdiagshort{\boldsymbol{0}, \ldots, \boldsymbol{0},\ &\tilde{\mH}_k,\ \boldsymbol{0}, \ldots, \boldsymbol{0}} \in \setC^{KN_t \times KN_t},\\
	    \bar{\mH}_k \triangleq \blkdiagshort{\tilde{\mH}_k, \ldots, \tilde{\mH}_k,\ &\boldsymbol{0},\ \tilde{\mH}_k, \ldots, \tilde{\mH}_k} \in \setC^{KN_t \times KN_t}.
	\end{align*}
	As a result, we can further rewrite constraint \eqref{cons_min_rate_BF2} as $ (\vw^H \hat{\mH}_k \vw)/ (\vw^H \bar{\mH}_k \vw + \sigma_k^2) \geq \gamma_{k}, \forall k$, which is equivalent to
	\begin{align*}
		 \vw^H \bar{\mH}_k \vw + \sigma_k^2 - \frac{\vw^H \hat{\mH}_k \vw}{\gamma_{k}} \leq 0,\ \forall k. \nbthis \label{cons_min_rate_BF2_1}
	\end{align*}
	 By applying the first-order Taylor approximation around  the point $[\vw^{(i)}, \gamma_k^{(i)}]$ found at iteration $i$, the concave function $f_{\mathtt{qol}}(\vw,\gamma_k) \triangleq - \vw^H \hat{\mH}_k \vw/\gamma_{k}$ is linearized as
	\begin{align*}
		f_{\mathtt{qol}}(\vw,\gamma_k) &\leq F_{\mathtt{qol}}(\vw,\gamma_k;\vw^{(i)}, \gamma_k^{(i)})\\
		&\triangleq \frac{\vw^{(i)H} \hat{\mH}_k \vw^{(i)}}{\gamma_k^{(i)2}} \gamma_k - \frac{2\re{\vw^{(i)H} \hat{\mH}_k \vw}}{\gamma_k^{(i)}}. \nbthis \label{eq_QOL}
	\end{align*}
	As a result, \eqref{cons_min_rate_BF2_1} can be transformed to the convex constraint
	\begin{align*}
		\vw^H \bar{\mH}_k \vw + \sigma_k^2 + F_{\mathtt{qol}}(\vw,\gamma_k;\vw^{(i)}, \gamma_k^{(i)}) \leq 0,\ \forall k. \nbthis \label{cons_min_rate_BF2_2}
	\end{align*}
	Given $\sum_{k=1}^{K} \norm{\vw_k}^2 = \norm{\vw}^2$, \eqref{cons_RIS_power} becomes
	\begin{align*}
		\sum_{n \in \setA} \abs{\an}^2 \left(\sigma_{\mathrm{r}}^2 +  \norm{\vh_{1,n}}^2 \norm{\vw}^2   \right) \leq \prismax. \nbthis \label{cons_ris_power_1}
	\end{align*}

%	we first rewrite $R_{k}$ in \eqref{eq_rate} as
%	\begin{align*}
%		R_{k} = \log \Big(\sum\nolimits_{j = 1}^K pi \abs{h_{j}}^2 + \sigma_j^2\Big) - \bar{r}_{k}, \nbthis \label{eq_sinr_22}
%	\end{align*}
%	where $\sigma_k^2 \triangleq \sigma_{\mathrm{r}}^2 \normshort{\hrr^H \bPsi}^2 + \sigma_{\mathrm{u}}^2$ and $\bar{r}_{k} \triangleq \log (\sum_{j \neq k}  \abs{h_j}^2 + \sigma_k^2 )$.	Now, constraint \eqref{cons_min_rate} can be rewritten as $\log \Big(\sum_{j = 1}^K pi \abs{h_{j}}^2 + \sigma_j^2\Big) \geq \tau +  \bar{r}_{k}, \forall k$, where $ \bar{r}_{k}$ is concave with $\{p_{j}\}_{j \neq k}$. By the first-order Taylor approximation around $\{^{(i)}\}_{j \neq k}$, an upper bound of $\bar{r}_{k}$ can be found  as
%	\begin{align*}
%		\bar{r}_{k} &\leq \bar{r}_{\mathrm{ub},k}^{(i)} \triangleq \log(\tilde{p}_k^{(i)}) + \frac{1}{\tilde{p}_k^{(i)}} \underset{{j \neq k}}{\sum} \abs{h_j}^2 ( - ^{(i)}). \nbthis \label{approx_Rub}
%	\end{align*}
%	where $\tilde{p}_k^{(i)} \triangleq \sum_{j \neq k} ^{(i)} \abs{h_j}^2 + \sigma_k^2$. As a result, \eqref{cons_min_rate} can be approximated as
%	\begin{align*}
%		& \log \left(p \abs{\vh_{k}}^2 + \sigma_k^2\right)  \geq \tau +  \bar{r}_{\mathrm{ub},k}^{(i)}, \forall k. \nbthis \label{cons_min_rate_approx}
%	\end{align*}
	Summary, we solve the following convex program of \eqref{problem_power} at iteration $i$:
	\begin{align}
		\label{problem_power_1}
		\underset{\tau,\vw}{\textrm{maximize}}\ \tau,\ \textrm{subject to}\ \eqref{cons_tx_power}, \eqref{cons_min_rate_BF1}, \eqref{cons_min_rate_BF2_2}, \eqref{cons_ris_power_1}.
	\end{align}

	\subsection{Optimization of RIS Coefficients}
	
	Given $\bW$, the RIS coefficients ($\ba$) can be optimized by solving the following problem
	\begin{align}
		\label{problem_alpha}
		\underset{\substack{\tau,\ba}}{\textrm{maximize}}\ \tau,\
		\textrm{subject to}\ \eqref{cons_min_rate}, \eqref{cons_phase}-\eqref{cons_RIS_power},
	\end{align}
	where constraints \eqref{cons_phase}-\eqref{cons_RIS_power} are convex with respect to $\ba$. The optimization variables $\ba$ have not been exposed in the current form of the nonconvex constraint \eqref{cons_min_rate}. To address this issue, we denote $\bar{h}_{0,kj} \triangleq \hd^H \vw_j$ and $\bar{\vh}_{1,k} \triangleq \mH_1 \vw_k, \forall k,j$,  yielding $\vh_{k}^H \vw_k = \bar{h}_{0,kk} + \hrr^H \bUpsilon \bar{\vh}_{1,k}$ and $\vh_{k}^H \vw_j = \bar{h}_{0,kj} + \hrr^H \bUpsilon \bar{\vh}_{1,j}$. Then, we can write the SINR term in \eqref{eq_rate} as
	\begin{align*}
		\mathrm{SINR} = \frac{\absshort{\bar{h}_{0,kk} + \hrr^H \bUpsilon \bar{\vh}_{1,k}}^2}{\sum\nolimits_{j \neq k} \absshort{\bar{h}_{0,kj} + \hrr^H \bUpsilon \bar{\vh}_{1,j}}^2 + \sigma_{\mathrm{r}}^2 \normshort{\hrr^H \bPsi}^2 + \sigma_{\mathrm{u}}^2}.
	\end{align*}
	By defining $\va \triangleq [\alpha_1, \ldots, \alpha_N ]^T \in \setC^{N \times 1}$, $\tilde{\mH}_{2,k} \triangleq \diagshort{\hrr^H}  \in \setC^{N \times N}$, and $\tilde{\vh}_{12,kj} \triangleq \tilde{\mH}_{2,k} \bar{\vh}_{1,j}  \in \setC^{N \times 1}$, we have $\bar{h}_{0,kj} + \hrr^H \bUpsilon \bar{\vh}_{1,j} = \bar{h}_{0,kj} + \va^T \tilde{\vh}_{12,kj}$ and $\hrr^H \bPsi = \va^T \mathbbm{1}^{\setA}_N \tilde{\mH}_{2,k}, \forall k, j$. After straightforward algebraic manipulations, the numerator and denominator of the SINR can be expressed as $\va^H \mQ_{k} \va + 2 \re{\va^H \vt_{k}} + e_{k}$ and $\va^H \tilde{\mQ}_{k} \va + 2\re{\va^H \tilde{\vt}_{k}} +  \tilde{e}_{k}$, respectively; Here, $\mQ_{k} = \tilde{\vh}_{12,kk}^* \tilde{\vh}_{12,kk}^T$, $\vt_{k} = \tilde{\vh}_{12,kk}^* \bar{h}_{0,kk}$, and $e_{k} = \abs{\bar{h}_{0,kk}}^2$; furthermore, $\tilde{\mQ}_{k} = \sigma_{\mathrm{r}}^2  \mathbbm{1}^{\setA}_N \tilde{\mH}_{2,k}^* \tilde{\mH}_{2,k}^T \mathbbm{1}^{\setA}_N + \sum_{j \neq k} \tilde{\vh}_{12,kj}^* \tilde{\vh}_{12,kj}^T$, $\tilde{\vt}_{k} = \sum_{j \neq k} \tilde{\vh}_{12,kj}^* \bar{h}_{0,kj}$, $\tilde{e}_{k} = \sigma_{\mathrm{u}}^2 + \sum_{j \neq k}  \abs{\bar{h}_{0,kj}}^2$, and $\re{\cdot}$ denotes the real part of a complex number. Thus, the rate function of UE $k$ is rewritten as
	\begin{align*}
		R_{k} = \log \left(1 + \frac{\va^H \mQ_{k} \va + 2 \re{\va^H \vt_{k}} + e_{k}}{ \va^H \tilde{\mQ}_{k} \va + 2\re{\va^H \tilde{\vt}_{k}} +  \tilde{e}_{k}}\right), \nbthis \label{eq_rate_alpha}
	\end{align*}
	where $\va$ is clearly exposed.
	By introducing new variables $\{\tilde{N}_k, \tilde{\gamma}_k\}$, constraint \eqref{cons_min_rate} is equivalently rewritten  as
	\begin{subequations}
		\begin{align*}
			\log(1 + \tilde{\gamma}_k) &\geq \tau,\ \forall k, \nbthis \label{cons_rate_1}\\
			\frac{\tilde{N}_k^2}{ \va^H \tilde{\mQ}_{k} \va + 2\re{\va^H \tilde{\vt}_{k}} +  \tilde{e}_{k}} &\geq \tilde{\gamma}_k,\ \forall k,  \nbthis \label{cons_gammatilde}\\
			\va^H \mQ_{k} \va + 2 \re{\va^H \vt_{k}} + e_{k} &\geq \tilde{N}_k^2,\ \forall k.  \nbthis \label{cons_rate_Ntilde}
		\end{align*}
	\end{subequations}
	The nonconvex constraints include \eqref{cons_gammatilde} and \eqref{cons_rate_Ntilde}, which are expressed as
	\begin{subequations}
		\begin{align*}
		\va^H \tilde{\mQ}_{k} \va + 2\re{\va^H \tilde{\vt}_{k}} +  \tilde{e}_{k} - \frac{\tilde{N}_k^2}{ \tilde{\gamma}_k } \leq 0,\ \forall k,  \nbthis \label{cons_gammatilde1}\\
		\tilde{N}_k^2 - \normshort{\bar{\mQ}_{k} \va}^2 - 2 \re{\va^H \vt_{k}} - e_{k} \leq 0,\ \forall k,  \nbthis \label{cons_rate_Ntilde1}
		\end{align*}
	\end{subequations}
	where $\bar{\mQ} = \mQ_k^{1/2}$. To address the nonconvexity of \eqref{cons_gammatilde1} and \eqref{cons_rate_Ntilde1}, we use the following approximations
	\begin{align*}
		-\frac{\tilde{N}_k^2}{ \tilde{\gamma}_k } &\leq F_{\mathtt{qol}}(\tilde{N}_k,\tilde{\gamma}_k;\tilde{N}_k^{(i)},\tilde{\gamma}_k^{(i)})\\
		%&\qquad \triangleq \left(\frac{\tilde{N}_k^{(i)}}{\tilde{\gamma}_k^{(i)}}\right)^2  \tilde{\gamma}_k - 2 \frac{\tilde{N}_k^{(i)}}{\tilde{\gamma}_k^{(i)}} \tilde{N}_k,\ \forall k, \nbthis \label{eq_Fqol} \\
		%%
		-\normshort{\bar{\mQ}_{k} \va}^2 &\leq F_{\mathtt{qua}}(\bar{\mQ}_{k} \va; \bar{\mQ}_{k} \va^{(i)}),\ \forall k, \nbthis \label{eq_Fqua}
	\end{align*}
	where $F_{\mathtt{qol}}(\cdot;\cdot)$ is defined in \eqref{eq_QOL}, and $F_{\mathtt{qua}}(\vx; \vx_0) \triangleq  2 \vx_0^H (\vx_0 - \vx) - \normshort{\vx_0}^2$ is a convex approximation of the concave function $-\norm{\vx}^2$ around $\vx_0$. As a result, \eqref{cons_gammatilde1} and \eqref{cons_rate_Ntilde1} are iteratively replaced by the following convex constraints
	\begin{subequations}
		\begin{align*}
			&\va^H \tilde{\mQ}_{k} \va + 2\re{\va^H \tilde{\vt}_{k}} +  \tilde{e}_{k}\\
			&\qquad + F_{\mathtt{qol}}(\tilde{N}_k,\tilde{\gamma}_k;\tilde{N}_k^{(i)},\tilde{\gamma}_k^{(i)}) \leq 0, \forall k,  \nbthis \label{cons_gammatilde2}\\
			&\tilde{N}_k^2 + F_{\mathtt{qua}}(\bar{\mQ}_{k} \va; \bar{\mQ}_{k} \va^{(i)}) - 2 \re{\va^H \vt_{k}} - e_{k} \leq 0.  \nbthis \label{cons_rate_Ntilde2}
		\end{align*}
	\end{subequations}
%	\begin{align}
%		\label{cons_min_rate_1}
%		\log \Big(\va^H \bar{\mQ}_{k} \va + 2 \re{\va^H \bar{\vt}_{k}} + \bar{e}_{k}\Big) \geq \tau + r_k,
%	\end{align}
%	where $\bar{\mQ}_{k} = \mQ_{k} + \tilde{\mQ}_{k}$, $\bar{\vt}_{k} = \vt_{k} + \tilde{\vt}_{k}$, $\bar{e}_k = e_k + \tilde{e}_k$, $r_k \triangleq \log( \va^H \tilde{\mQ}_{k} \va +2\re{\va^H \tilde{\vt}_{k}} +  \tilde{e}_{k})$, and $r_k$ is concave with respect to $\va$. We introduce slack variables $b_{k}$ which satisfies $b_{k} \geq \va^H \tilde{\mQ}_{k} \va +2\re{\va^H \tilde{\vt}_{k}}$. With this introduction of $b_{k}$, we can write $r_k = \log(b_{k} + \tilde{e}_{k})$. Thus, an upper bound of $r_k$ can be found as
%	\begin{align*}
%		r_k \leq r_{\mathrm{ub},k}^{(i)} \triangleq \log\Big( b_{k}^{(i)} + \tilde{e}_{k} \Big) + \frac{b_{k} - b_{k}^{(i)}}{b_{k}^{(i)} + \tilde{e}_{k}}, \nbthis \label{rate_UB}
%	\end{align*}
%	based on the first order Taylor approximation around $b_{k}^{(i)}$. As a result, constraint \eqref{cons_min_rate_1} can be approximated by the following set of convex constraints:
%	\begin{subequations}
%		\begin{align}
%			\log (\va^H \bar{\mQ}_{k} \va + 2 \re{\va^H \bar{\vt}_{k}} + e_{k}) \geq \tau + r_{\mathrm{ub},k}^{(i)},\ \forall k \nbthis \label{cons_min_rate_2a} \\
%			b_{k} \geq \va^H \tilde{\mQ}_{k} \va + 2\re{\va^H \tilde{\vt}_{k}},\ \forall k. \nbthis \label{cons_min_rate_2b}
%		\end{align}
%	\end{subequations}
	Furthermore,  \eqref{cons_RIS_power} is transformed to a more compact form as
	\begin{align*}
		\va^H \bXi \va \leq \prismax, \nbthis \label{cons_RIS_power2}
	\end{align*}
	where $\bXi = \diagshort{\tilde{\xi}_1, \ldots, \tilde{\xi}_N}$ with $\tilde{\xi}_n = \xi_n$ for $n \in \setA$, and $\tilde{\xi}_n = 0$, otherwise. 
	
	Finally, problem \eqref{problem_alpha} can be approximated by the following convex program at iteration $i$
	\begin{align}
		\label{problem_alpha_1}
		\hspace{-0.25cm}\underset{\substack{\tau,\va,\{\tilde{N}_{k}\},\{\tilde{\gamma}_{k}\}}}{\textrm{maximize}}\ \tau,\
		\textrm{s. t.}\  \eqref{cons_phase}-\eqref{cons_active_modul}, \eqref{cons_rate_1}, \eqref{cons_gammatilde2}-\eqref{cons_RIS_power2}.
	\end{align}

	%\subsection{Overall Algorithm}
	
	We summarize the proposed iterative algorithm based on the BCA and SCA methods for solving \eqref{prob_0} in Algorithm \ref{alg_opt}. In step 1, the initial points for $\{\vw_k^{(0)}\}$, $\{\alpha_n^{(0)}\}$, $\{\tilde{N}_{k}^{(0)}\}$, and $\{\tilde{\gamma}_{k}^{(0)}\}$ are generated to guarantee that Algorithm \ref{alg_opt} is successfully solved in the first iteration. In steps 2--6, subproblems \eqref{problem_power_1} and \eqref{problem_alpha_1} are alternatively solved and $\{\vw_k^{(i)}\}$, $\{\alpha_n^{(i)}\}$, $\{\tilde{N}_{k}^{(i)}\}$, and $\{\tilde{\gamma}_{k}^{(i)}\}$ are updated after each iteration until the objective value $\tau$ converges. Note that convex problems \eqref{problem_power_1} and \eqref{problem_alpha_1} can be solved with standard optimization toolbox, such as CVX or YALMIP-MOSEK. The computational complexities required to solve subproblems \eqref{problem_power_1} and \eqref{problem_alpha_1} are $\mathcal{O}\left( \sqrt{2K+2} (KN_t + 1)^3 \right)$ and $\mathcal{O}\left( \sqrt{3K+N+1} (2K+N+1)^3 \right)$, respectively. Thus, the overall complexity of Algorithm \ref{alg_opt} is $\mathcal{O}\left(I (\sqrt{2K+2} (KN_t + 1)^3 + \sqrt{3K+N+1} (2K+N+1)^3) \right)$, where $I$ is the number of iterations until convergence.
	
	\begin{algorithm}[t]
		\small
		\caption{Iterative Algorithm to Solve Problem \eqref{prob_0}}
		\label{alg_opt}
		\begin{algorithmic}[1]
			\STATE \textbf{Initialize} $\{\vw_k^{(0)}, \alpha_n^{(0)}\}$. The feasible points for $\{\tilde{N}_{k}^{(0)},\tilde{\gamma}_{k}^{(0)}\}$ are set to hold the equalities in \eqref{cons_gammatilde} and \eqref{cons_rate_Ntilde}. Set $i=0$. 
			
			\REPEAT

			\STATE Solve problem \eqref{problem_power_1} for given $\{\alpha_n^{(i)}\}$ to obtain the solution $\{\vw_k^{\star}\}$. Update $ \{\vw_k^{(i+1)}\} := \{\vw_k^{\star}\}$.
			
			\STATE Solve problem \eqref{problem_alpha_1} for given $\{\vw_k^{(i+1)}\}$ to obtain the solutions $\va^{\star}, \{\tilde{N}_{k}^{\star}\},\{\tilde{\gamma}_{k}^{\star}\}$. Update $\an^{(i+1)} := \an^{\star}, \forall n$,  $\tilde{N}_{k}^{(i+1)} := \tilde{N}_{k}^{\star}$, and $\tilde{\gamma}_{k}^{(i+1)} := \tilde{\gamma}_{k}^{\star}, \forall k$.

			\STATE Set $i=i+1$.
			\UNTIL convergence.
			%\STATE Set $\an = \frac{\an}{\abs{\an}}, \forall n \in \setA$.
		\end{algorithmic}
	\end{algorithm}
	
	\section{Numerical Results}
	
	In this section, numerical results are provided to evaluate the effectiveness of Algorithm \ref{alg_opt}. We assume that the BS and RIS are deployed in a two-dimensional coordinate system at $(0,0)$ and $(20,0)$ m, respectively, while the UEs are randomly and uniformly distributed in a square area of $200 \times 200$ $\text{m}^2$. The Rayleigh fading model is considered for the direct BS-UEs channels, while those for the BS-RIS and RIS-UEs reflecting channels are Rician fading models with Rician factors of $100$ and $10$, respectively. The path loss for link distance $d$ is given by $\beta(d) = \beta_0 (d/1\mathrm{ m} )^{-\epsilon}$, where $\beta_0$ is the path loss at the reference distance of $1$ m, and $\epsilon \in \{\epsilon_0, \epsilon_1, \epsilon_2\}$ represents the path loss exponents of BS-UEs, BS-RIS, and RIS-UEs channels, respectively. We set $\beta_0 = -30$ dB and $\sigma^2_{\mathrm{u}} = -80$ dBm, $\{\epsilon_0, \epsilon_1, \epsilon_2\} = \{3.2,2.2,2.5\}$. The total power of noise and residual self-interference of the RIS is computed as $\sigma^2_{\mathrm{r}} = (\eta + 1) \sigma^2_{\mathrm{u}}$, with $\eta = 1$ dB reflecting the possible residual self-interference caused by active elements operating  in full-duplex mode \cite{malik2018optimal, nguyen2021hybrid}. The positions of the RIS active elements are fixed to $\setA = \{1,\ldots,\Na\}$. Towards a fair comparison, the power budget at the BS in the hybrid RIS-aided system is reduced by $\prismax$, so that all the compared schemes have the same total power budget as $\ptmax$.% Other simulation parameters are shown in the captions of the figures.
	
	%\subsection{Convergence of Algorithm \ref{alg_opt}}
	\begin{figure}[t]
		%\centering
		\vspace{-0.5cm}
		%\belowcaptionskip = -15pt
		\includegraphics[scale=0.6]{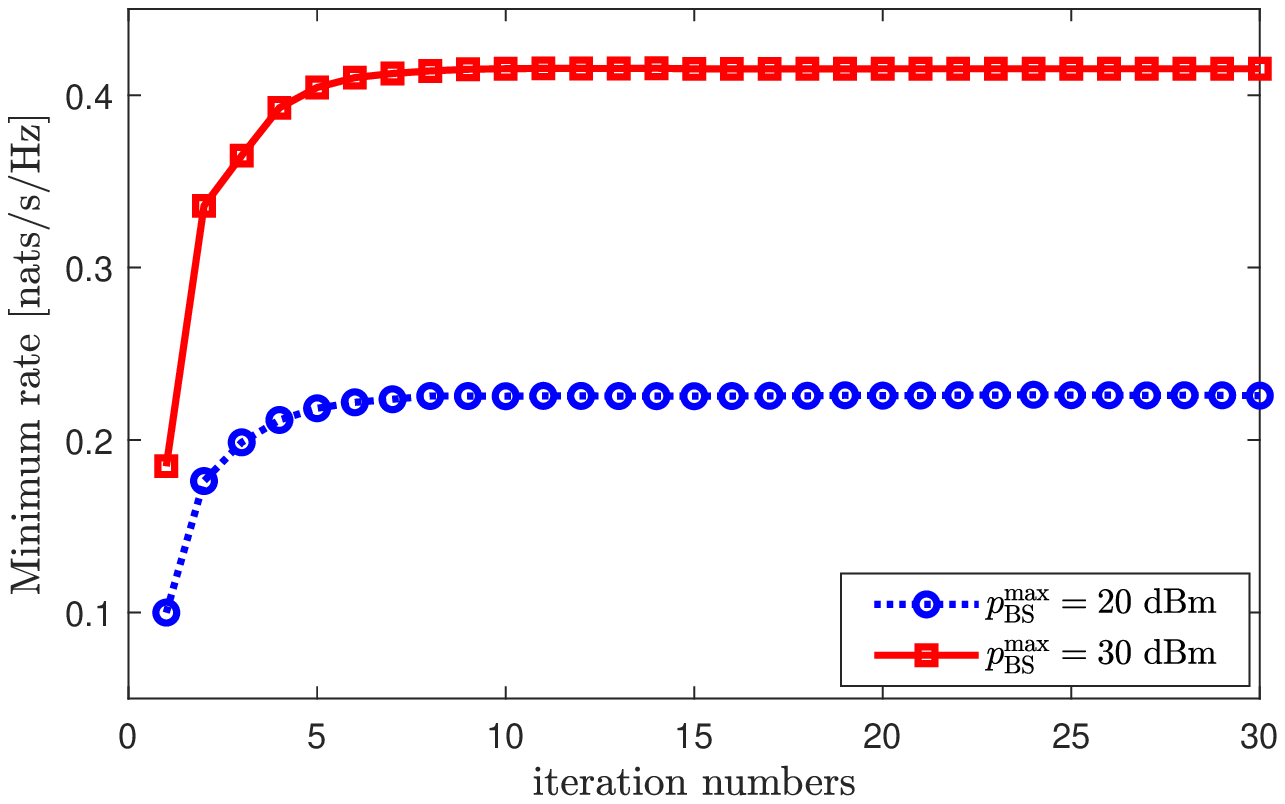}
		\caption{Convergence of Algorithm \ref{alg_opt} with $N_t=2$, $K=5$, $N = 50$, $\ptmax=\{20,30\}$ dBm, $\Na=4$ and $\prismax = 0$ dBm.}
		\label{fig_rate_conv}
	\end{figure}
	
	We first show in Fig.\ \ref{fig_rate_conv} the convergence of Algorithm \ref{alg_opt}. For initialization, we set $\vw_k^{(0)} = \sqrt{\frac{\ptmax}{K}} \frac{\vh_k}{\normshort{\vh_k}}, \forall k$, implying the conjugate beamforming with equal power. Furthermore, $\{\alpha_n^{(0)}\}$ are initialized as $\{r e^{j \theta_n^{(0)}}\}$ with $\{\theta_n^{(0)}\}$ being randomly generated on $[0, 2\pi)$ and $0 < r \leq \frac{\prismax}{\sum_{n \in \setA} \xi_n}$. These initial values belong to the feasible region of problem \eqref{prob_0}. For both cases $\ptmax=\{20,30\}$ dBm, it is observed  that Algorithm \ref{alg_opt} converges after only a few iterations. With $\ptmax=20$ dBm, the algorithm converges slightly faster, but obviously to a lower rate, compared with the case $\ptmax=30$ dBm.% Furthermore, it is also clear that the hybrid RIS scheme converges to higher objective values in both scenarios.
	
	%\subsection{Performance Improvement of the Hybrid RIS}
	
	\begin{figure}[t]
		%\hspace{-0.5cm}
		%\belowcaptionskip = -15pt
		\includegraphics[scale=0.6]{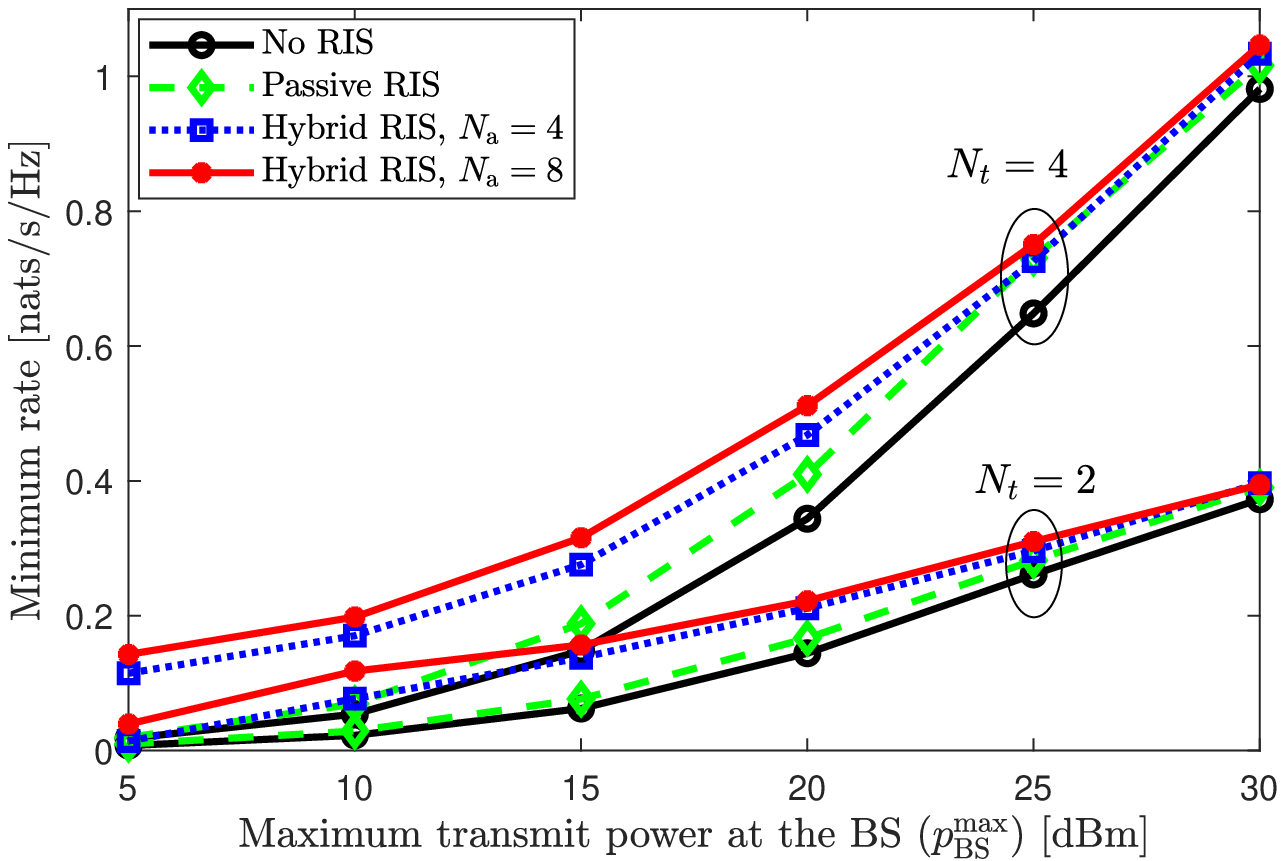}
		\caption{Minimum rates versus $\ptmax$ with $N_t=\{2,4\}$, $K=5$, $N = 50$, $\Na = \{4,8\}$, $\ptmax=[0,30]$ dBm, and $\prismax = -1$ dBm.}
		\label{fig_rate_vs_Pt}
	\end{figure}
	
	In Fig.\ \ref{fig_rate_vs_Pt}, we plot the minimum rate versus the maximum transmit power of the BS with various deployments of RISs. As expected, the minimum rate performance of both  passive and hybrid RISs increases significantly when $\ptmax$ increases for both $N_t = \{2,4\}$. However, the performance gain of the former is marginal, especially at low SNRs. In contrast, the hybrid RIS with only $\Na=4$ active elements can provide significant performance improvement to the system, and the gain is more remarkable at a low SNR regime. For example, with $N_t = 4$ and $\ptmax = 20$ dBm, the passive RIS achieves $27 \%$ improvement, while that attained by the hybrid RIS is up to $80 \%$ with the requirement of only $\Na = 4$ active elements and a power budget of $\prismax = -1$ dBm. Furthermore, the hybrid RIS with $\Na = 8$ provides only slightly better performance compared to the case $\Na = 4$.%, as will be further discussed below.

	\begin{figure}[t]
		%\centering
		\vspace{-0.25cm}
		%\belowcaptionskip = -15pt
		\includegraphics[scale=0.6]{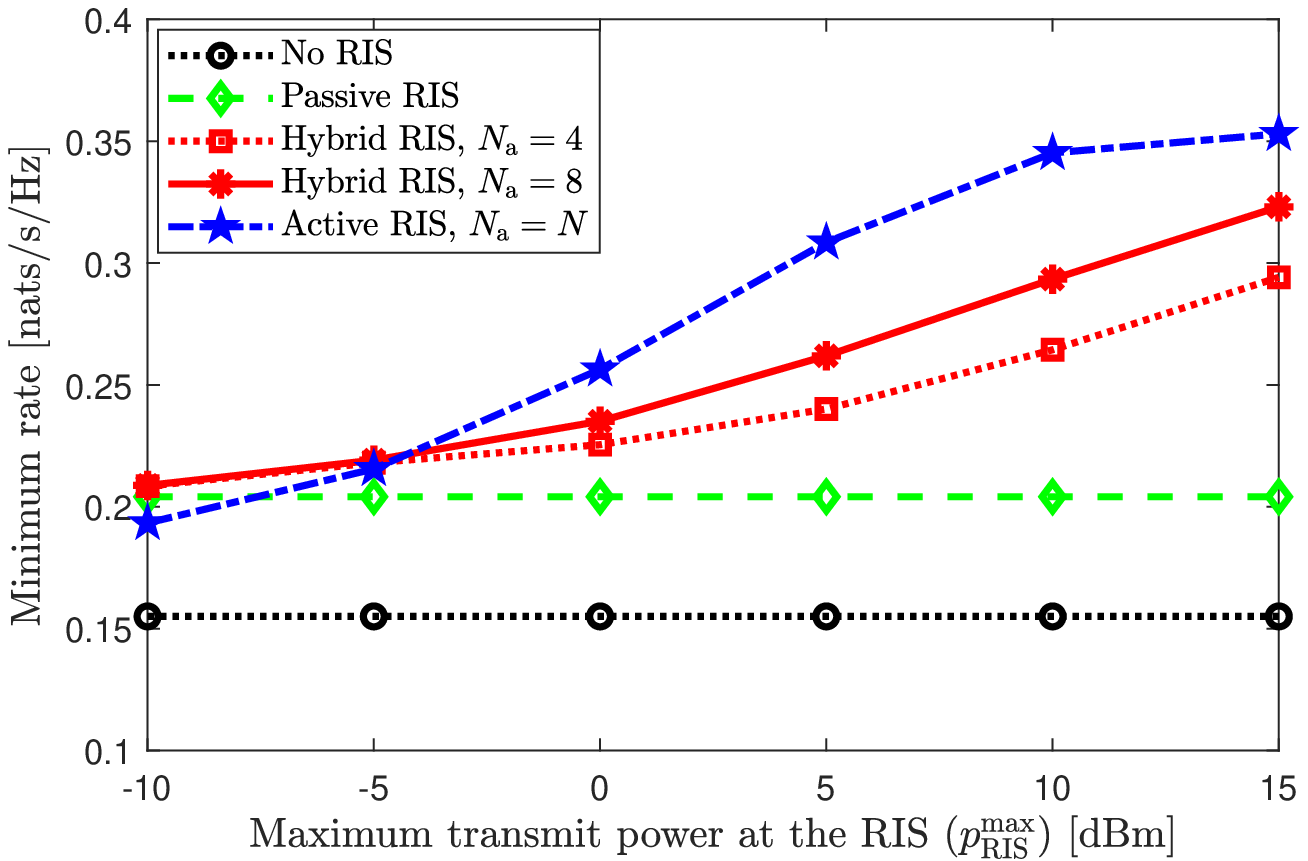}
		\caption{Minimum rates versus $\prismax$ with $N_t=2$, $K=5$, $N = 50$, $\Na = \{4,8,N\}$, $\ptmax=20$ dBm, and $\prismax = [-10,15]$ dBm.}
		\label{fig_rate_vs_Pris}
	\end{figure}
	
	In Fig.\ \ref{fig_rate_vs_Pris}, we show the minimum rate versus the transmit power budget at RISs, i.e., $\prismax$. Unsurprisingly, the performance improvement attained by the hybrid RIS significantly increases with $\prismax$. Furthermore, as $\prismax$ increases, the hybrid RIS with more active elements provides significantly larger performance gains. However, this is not valid for a low $\prismax$. It is observed that  hybrid RISs with $\Na = 4$ and $\Na = 8$ offer almost the same performance for $\prismax \leq -5$ dBm, while the fully active RIS (with $\Na = N = 50$) performs even worse than the passive RIS. This is because when a limited power budget needs to be shared among a large number of active elements, the amplitudes of these elements become very small, causing signal attenuation on reflecting channels. Note that a RIS with a larger number of active elements also requires a higher power consumption \cite{long2021active}. These further explain the motivations of hybrid RISs, especially when the power budget allocated to RISs is limited.% Specifically, a hybrid RIS with appropriate number of active elements can guarantee performance improvement even with a limited power budget.

	\section{Conclusion}
	
	We proposed the deployment of the hybrid active-passive RIS architecture to enhance the performance of multi-user MISO systems. In the hybrid RIS, very few active elements are employed to enhance reflecting and amplifying gains of the RISs' elements. We proposed an iterative algorithm based on the BCA and SCA approaches to effectively solve the formulated max-min rate problem. Numerical results were provided to verify the merits of the proposed algorithm. They revealed that the hybrid RIS offers remarkable performance improvement compared to existing schemes (i.e. without RIS and with conventional passive RISs).% demonstrating the potential benefits of the hybrid RIS in the considered system. %The applied channel models and the assumption of MISO links make the results relevant to the conventional microwave frequencies (below 10 GHz) demonstrating the potential benefits of the HR-RIS also in that regime, while the majority of RIS research is directed towards the mmWave bands.
	
	%\section*{Acknowledgment}
	%This work has been supported in part by Academy of Finland under 6Genesis Flagship (grant 318927), EERA Project (grant 332362), and Infotech Program funded by University of Oulu Graduate School.

	%\begingroup
    %\setstretch{0.95}
	\bibliographystyle{IEEEtran}
	\bibliography{IEEEabrv,Bibliography}

\begin{thebibliography}{10}
\providecommand{\url}[1]{#1}
\csname url@samestyle\endcsname
\providecommand{\newblock}{\relax}
\providecommand{\bibinfo}[2]{#2}
\providecommand{\BIBentrySTDinterwordspacing}{\spaceskip=0pt\relax}
\providecommand{\BIBentryALTinterwordstretchfactor}{4}
\providecommand{\BIBentryALTinterwordspacing}{\spaceskip=\fontdimen2\font plus
\BIBentryALTinterwordstretchfactor\fontdimen3\font minus
  \fontdimen4\font\relax}
\providecommand{\BIBforeignlanguage}[2]{{%
\expandafter\ifx\csname l@#1\endcsname\relax
\typeout{** WARNING: IEEEtran.bst: No hyphenation pattern has been}%
\typeout{** loaded for the language `#1'. Using the pattern for}%
\typeout{** the default language instead.}%
\else
\language=\csname l@#1\endcsname
\fi
#2}}
\providecommand{\BIBdecl}{\relax}
\BIBdecl

\bibitem{huang2019reconfigurable}
C.~Huang, A.~Zappone, G.~C. Alexandropoulos, M.~Debbah, and C.~Yuen,
  ``Reconfigurable intelligent surfaces for energy efficiency in wireless
  communication,'' \emph{{IEEE} Trans. Wireless Commun.}, vol.~18, no.~8, pp.
  4157--4170, 2019.

\bibitem{wu2019intelligent}
Q.~Wu and R.~Zhang, ``Intelligent reflecting surface enhanced wireless network
  via joint active and passive beamforming,'' \emph{{IEEE} Trans. Wireless
  Commun.}, vol.~18, no.~11, pp. 5394--5409, 2019.

\bibitem{wu2019towards}
------, ``Towards smart and reconfigurable environment: {I}ntelligent
  reflecting surface aided wireless network,'' \emph{{IEEE} Commun. Mag.},
  vol.~58, no.~1, pp. 106--112, 2019.

\bibitem{hu2020reconfigurable}
J.~Hu, Y.-C. Liang, and Y.~Pei, ``Reconfigurable intelligent surface enhanced
  multi-user {MISO} symbiotic radio system,'' \emph{{IEEE} Trans. Commun.},
  vol.~69, no.~4, pp. 2359--2371, 2020.

\bibitem{di2020hybrid}
B.~Di, H.~Zhang, L.~Song, Y.~Li, Z.~Han, and H.~V. Poor, ``Hybrid beamforming
  for reconfigurable intelligent surface based multi-user communications:
  {A}chievable rates with limited discrete phase shifts,'' \emph{{IEEE} J. Sel.
  Areas Commun.}, vol.~38, no.~8, pp. 1809--1822, 2020.

\bibitem{kammoun2020asymptotic}
A.~Kammoun, A.~Chaaban, M.~Debbah, M.-S. Alouini \emph{et~al.}, ``Asymptotic
  max-min {SINR} analysis of reconfigurable intelligent surface assisted {MISO}
  systems,'' \emph{{IEEE} Trans. Wireless Commun.}, vol.~19, no.~12, pp.
  7748--7764, 2020.

\bibitem{zhang2021reconfigurable}
H.~Zhang, B.~Di, Z.~Han, H.~V. Poor, and L.~Song, ``Reconfigurable intelligent
  surface assisted multi-user communications: {H}ow many reflective elements do
  we need?'' \emph{{IEEE} Wireless Commun. Lett.}, vol.~10, no.~5, pp.
  1098--1102, 2021.

\bibitem{li2020interference}
A.~Li, L.~Song, B.~Vucetic, and Y.~Li, ``Interference exploitation precoding
  for reconfigurable intelligent surface aided multi-user communications with
  direct links,'' \emph{{IEEE} Wireless Commun. Lett.}, vol.~9, no.~11, pp.
  1937--1941, 2020.

\bibitem{gan2021ris}
X.~Gan, C.~Zhong, C.~Huang, and Z.~Zhang, ``Ris-assisted multi-user {MISO}
  communications exploiting statistical {CSI},'' \emph{{IEEE} Trans. Wireless
  Commun.}, vol.~69, no.~10, pp. 6781--6792, 2021.

\bibitem{di2020practical}
B.~Di, H.~Zhang, L.~Li, L.~Song, Y.~Li, and Z.~Han, ``Practical hybrid
  beamforming with finite-resolution phase shifters for reconfigurable
  intelligent surface based multi-user communications,'' \emph{{IEEE} Trans.
  Veh. Technol.}, vol.~69, no.~4, pp. 4565--4570, 2020.

\bibitem{huang2020reconfigurable}
C.~Huang, R.~Mo, and C.~Yuen, ``Reconfigurable intelligent surface assisted
  multiuser {MISO} systems exploiting deep reinforcement learning,''
  \emph{{IEEE} J. Sel. Areas Commun.}, vol.~38, no.~8, pp. 1839--1850, 2020.

\bibitem{ma2021joint}
X.~Ma, S.~Guo, H.~Zhang, Y.~Fang, and D.~Yuan, ``Joint beamforming and
  reflecting design in reconfigurable intelligent surface-aided multi-user
  communication systems,'' \emph{{IEEE} Trans. Wireless Commun.}, vol.~20,
  no.~5, pp. 3269--3283, 2021.

\bibitem{huang2018energy}
C.~Huang, G.~C. Alexandropoulos, A.~Zappone, M.~Debbah, and C.~Yuen, ``Energy
  efficient multi-user {MISO} communication using low resolution large
  intelligent surfaces,'' in \emph{IEEE Global Commun. Conf. (GLOBECOM)
  Workshop}, 2018, pp. 1--6.

\bibitem{taha2019deep}
A.~Taha, M.~Alrabeiah, and A.~Alkhateeb, ``{Deep learning for large intelligent
  surfaces in millimeter wave and massive MIMO systems},'' in \emph{IEEE Global
  Commun. Conf. (GLOBECOM)}, 2019, pp. 1--6.

\bibitem{alexandropoulos2020hardware}
G.~C. Alexandropoulos and E.~Vlachos, ``A hardware architecture for
  reconfigurable intelligent surfaces with minimal active elements for explicit
  channel estimation,'' in \emph{Proc. IEEE Int. Conf. Acoust., Speech, Signal
  Processing}, 2020, pp. 9175--9179.

\bibitem{nguyen2021spectral}
N.~T. Nguyen, Q.-D. Vu, K.~Lee, and M.~Juntti, ``Spectral efficiency
  optimization for hybrid relay-reflecting intelligent surface,'' \emph{Proc.
  IEEE Int. Conf. Commun. Workshop}, 2021.

\bibitem{nguyen2021hybrid}
------, ``{Hybrid relay-reflecting intelligent surface-assisted wireless
  communication},'' \emph{to appear in IEEE Trans. Veh. Technol.}, 2022.

\bibitem{nguyen2021hybrid_mag}
{N. T. Nguyen \textit{et al.}}, ``Hybrid relay-reflecting intelligent
  surface-aided wireless communications: Opportunities, challenges, and future
  perspectives,'' \emph{arXiv preprint arXiv:2104.02039}, 2021.

\bibitem{nguyen2022downlink}
------, ``Downlink throughput of cell-free massive {MIMO} systems assisted by
  hybrid relay-reflecting intelligent surfaces,'' in \emph{Proc. IEEE Int.
  Conf. Commun.}, 2022.

\bibitem{zeng2021active}
P.~Zeng, D.~Qiao, Q.~Wu, and Y.~Wu, ``Active {IRS} aided {WPCNs}: {A} new
  paradigm towards higher efficiency and wider coverage,'' \emph{arXiv preprint
  arXiv:2111.11600}, 2021.

\bibitem{9598322}
{K.-H. Ngo \textit{et al.}}, ``Low-latency and secure computation offloading
  assisted by hybrid relay-reflecting intelligent surface,'' in \emph{IEEE Int.
  Conf. Advanced Tech. Commun. (ATC)}, 2021, pp. 306--311.

\bibitem{9653007}
S.~Ahmed, A.~E. Kamal, and M.~Y. Selim, ``Adding active elements to
  reconfigurable intelligent surfaces to enhance energy harvesting for {IoT}
  devices,'' in \emph{IEEE Military Commun. Conf.}, 2021, pp. 297--302.

\bibitem{DBLP:journals/corr/abs-2111-08355}
Z.~Yigit, E.~Basar, M.~Wen, and I.~Altunbas, ``Hybrid reflection modulation,''
  \emph{CoRR}, vol. abs/2111.08355, 2021.

\bibitem{long2021active}
R.~Long, Y.-C. Liang, Y.~Pei, and E.~G. Larsson, ``Active reconfigurable
  intelligent surface aided wireless communications,'' \emph{{IEEE} Trans.
  Wireless Commun.}, 2021.

\bibitem{khoshafa2021active}
M.~H. Khoshafa, T.~M. Ngatched, M.~H. Ahmed, and A.~R. Ndjiongue, ``Active
  reconfigurable intelligent surfaces-aided wireless communication system,''
  \emph{{IEEE} Commun. Lett.}, vol.~25, no.~11, pp. 3699--3703, 2021.

\bibitem{chen2022active}
G.~Chen, Q.~Wu, C.~He, W.~Chen, J.~Tang, and S.~Jin, ``Active {IRS} aided
  multiple access for energy-constrained {IoT} systems,'' \emph{arXiv preprint
  arXiv:2201.12565}, 2022.

\bibitem{schroeder2020passive}
R.~Schroeder, J.~He, and M.~Juntti, ``Passive {RIS} vs. {H}ybrid {RIS}: {A}
  comparative study on channel estimation,'' in \emph{Proc. IEEE Veh. Technol.
  Conf.}, June 2021, pp. 1--7.

\bibitem{landsberg2017low}
N.~Landsberg and E.~Socher, ``{A low-power 28-nm CMOS FD-SOI reflection
  amplifier for an active F-band reflectarray},'' \emph{IEEE Trans. Microw.
  Theory Techn.}, vol.~65, no.~10, pp. 3910--3921, 2017.

\bibitem{malik2018optimal}
R.~Malik and M.~Vu, ``Optimal transmission using a self-sustained relay in a
  full-duplex {MIMO} system,'' \emph{{IEEE} J. Sel. Areas Commun.}, vol.~37,
  no.~2, pp. 374--390, 2018.

\end{thebibliography}
	%\endgroup
\end{document}